\documentclass[preprint]{iucr}              
     \journalcode{S}              

\papertype{FA}
\usepackage{gensymb}
\usepackage[]{textcomp}
\usepackage{amsmath}
\usepackage{amstext}
\usepackage[final]{changes}
\usepackage[version=4]{mhchem}
\usepackage{verbatim}
\newcommand{\um}{$\mu$m }

\newcommand{\qz}{$\vec{q}_\text{z}$}

\begin{document}                  
\title{Quick X-ray Reflectivity using Monochromatic Synchrotron Radiation for Time-Resolved Applications}


 \cauthor[a,b]{H.}{Joress}{hj335@cornell.edu}{}
          \author[a,c]{J.D.}{Brock,}
          \author[a]{A.R.}{Woll}

     \aff[a]{Cornell High Energy Synchrotron Source, Cornell University \city{Ithaca}, NY, \country{United States}}
     \aff[b]{Materials Science and Engineering Department, Cornell University \city{Ithaca}, NY, \country{United States}}
     \aff[c]{School of Applied \& Engineering Physics, Cornell University \city{Ithaca}, NY, \country{United States}}


\shortauthor{Joress, Woll and Brock}


%
%



\maketitle                        


\begin{abstract}
We describe and demonstrate a new technique for parallel collection of x-ray reflectivity data, compatible with monochromatic synchrotron radiation and flat substrates, and apply it to the in-situ observation of thin-film growth. 
The method employs a polycapillary x-ray optic to produce a converging fan of radiation, incident onto a sample surface, and an area detector to simultaneously collect the XRR signal over an angular range matching that of the incident fan. Factors determining the range and instrumental resolution of the technique in reciprocal space, in addition to the signal-to-background ratio, are described in detail. 
Our particular implementation records $\sim$5\degree{} in $2\theta$ and resolves Kiessig fringes from samples with layer thicknesses ranging from 3 to 76  nm. 
We illustrate the value of this approach by showing in-situ XRR data obtained with 100 ms time resolution during the growth of epitaxial \ce{La_{0.7}Sr_{0.3}MnO3} on \ce{SrTiO3} by Pulsed Laser Deposition (PLD) at the Cornell High Energy Synchrotron Source (CHESS).
Compared to prior methods for parallel XRR data collection, ours is the first method that is both sample-independent and compatible with the highly collimated, monochromatic radiation typical of 3rd generation synchrotron sources. Further, our technique can be readily adapted for use with laboratory-based sources. 
%
%

\end{abstract}

\section{Introduction}
\label{sect:intro}

Specular X-ray Reflectivity (XRR) is a decades-old and well-established application of elastic x-ray scattering used to characterize surfaces and thin films, allowing for the determination of the thickness of films, the periodicity of multilayers,  and the roughness of interfaces \cite{kiessig1931interferenz,Parratt}. In general, elastic x-ray scattering is performed by measuring the x-ray intensity as a function of the scattering vector, $\vec{q}$, defined as the vector difference of the outgoing and incident wavevectors:  $\vec{k}_\text{out}-\vec{k}_\text{in}$.  XRR, in particular, is defined by the constraint that the incident and exit angles are equal and that $\vec{q}$ is parallel to the surface normal, $\hat{n}$. A key strength of XRR is that intensity modulations along \qz{}, the component of $\vec{q}$ along $\hat{n}$, can be very accurately calculated from a knowledge of $\rho(z)$, the projection of electron density along $\hat{n}$. In practice, XRR is used to obtain $\rho(z)$ from the sub-angstrom to the micron scale.

In addition to its use for routine characterization of static samples, XRR and other variants of surface-sensitive x-ray scattering methods \textemdash{}  especially grazing incidence diffraction (GID)  and grazing incidence small angle scattering (GISAXS) \textemdash{}  are well suited to the study of in-situ processes such as thin-film growth \cite{kowarik_rev}. This advantage arises, first, because x-ray diffraction is a remote probe; neither the source nor  the detector need be near the sample. Second, the weakly interacting nature of x-rays makes them compatible with a variety of sample environments, such as ambient pressure or liquids, not suitable for electrons and many other probes.
Recent examples of XRR for in-situ processes include  studies of pulsed laser deposition (PLD)\cite{Ferguson,bauer,chinta_sputtering}, molecular beam epitaxy (MBE)\cite{Nahm,lee_dynamic}, atomic layer deposition (ALD)\cite{Devloo,Klug,ju2017role}, sputter deposition \cite{Krause,sinsheimer2013situ,bein2015situ}, and electro-chemical reactions \cite{chang,Golks,Plaza}. 

The most common method for performing XRR measurements, originally developed by \citeasnoun{Parratt}, employs collimated, monochromatic radiation as the incident beam. Scattered intensity at different points along $\vec{q} \parallel \hat{n}$ are obtained by varying the incident angle, $\theta_\text{in}$, and exit angle, $\theta_\text{out}$, (both measured from the sample surface) while maintaining the specular condition $\theta_\text{in}=\theta_\text{out}$. The minimum and maximum length scales that can be characterized are determined by the angular range and resolution of the measurement, respectively. 

A disadvantage of the Parratt method for studying dynamic processes is that the intensities at each value of \qz{} are obtained sequentially, rather than in parallel. 
The necessity that the sample, and in some cases the detector, must be rotated relative to the source to change $|\vec{q}|$ while maintaining  $\vec{q} \parallel \hat{n}$ often limits the time resolution of the technique.  Methods have been developed to perform this scanning rapidly with scanning times as low as 2 seconds \cite{lippmann2016new,bein2015situ,mocuta2018fast}, however these methods are still limited by mechanical speed. This limitation can be particularly severe when rotating the sample requires moving a large or complex sample chamber, such as a thin-film deposition system.  

In order to improve the time resolution of XRR data collection, alternative methods involving parallel data collection have been developed. These methods, broadly described as quick XRR (qXRR), were recently reviewed by \citeasnoun{sakurai}. The oldest of these, developed by \citeasnoun{bilderback} utilizes white beam and an energy-dispersive point detector placed in the specular condition. The incident and exit angle are held fixed. The scattered intensity measured at different energies correspond to different magnitudes of $\vec{k}_\text{in}$ and $\vec{k}_\text{out}$ and so different values of \qz{}. This approach has been successfully implemented using both lab-based sources \cite{windover_edxrr,albertini} and synchrotron radiation \cite{weber_edxrr,kowarik2007energy} to monitor thin film growth. A challenge of this approach using lab sources is the absence of a bright, large-bandwidth beam. At synchrotrons, while both bend-magnet and wiggler sources provide nearly uniform intensity over a large energy range, it is difficult to  limit the bandwidth to match the needs of a particular experiment. Since XRR data can vary by several orders of magnitude over a relatively small \qz{} range, a more significant limitation is the limited dynamic range of energy-dispersive detectors. As a result, weak portions of the spectrum can be severely count-rate limited even when the detector is nearly saturated by contributions from bright regions of the XRR signal. 

A notable variation to the energy dispersive approach to qXRR, recently described and demonstrated in \citeasnoun{matsushita2008high} and \citeasnoun{japcurvedxtal}, overcomes two of the challenges described above (control of the incident-beam bandwidth and dynamic range limitations of the detector) using a 
combination of a bent Laue monochromator and area detector. In this geometry, the monochromator is used to create a converging, dispersive  fan of radiation on the sample. The beam and sample are oriented such that all rays in the fan have the same incident angle relative to the sample surface, but strike the surface at a range of different azimuthal angles around the surface normal. Because these rays vary in energy and hence wave-vector magnitude $|\vec{k}_{in}|$, the specularly reflected intensity at different points along \qz{} appear along a line on the detector parallel to the sample surface.   

In addition to approaches to qXRR using polychromatic radiation, several methods employ monochromatic radiation. These methods work by simultaneously illuminating the sample with a range of incident angles. The earliest example of this approach, developed by \citeasnoun{herbette1977direct} and implemented elsewhere  \cite{bosio1989situ,Liu:rx5034}, achieves this by using curved samples and a wide parallel beam.  Because of the shape of the surface, the incident beam will impinge on the sample with a range of angles across the width of the beam.  The diffracted intensity can than be recorded on a 1D or 2D detector with each point along the detector having a specular reflection with a unique \qz{}.  In this case the radius of curvature  determines the range of incident angles for a given incident beam width.

\begin{figure}
\caption{Schematic representation of angular dispersive, monochromatic qXRR. A monochromatic fan of radiation is incident on the surface. Different positions on an area detector, at right, collect different \qz{}-points along the XRR trajectory. The red, green, and purple lines are specific incident vectors and their associated specular reflections.}
\includegraphics{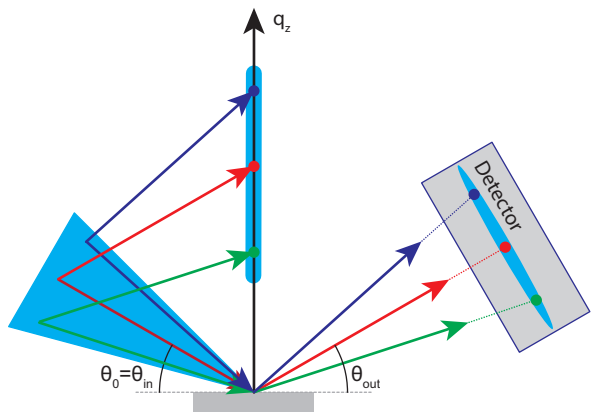}
\label{recipmap}
\end{figure}

A second approach to monochromatic qXRR, illustrated in Fig. \ref{recipmap}, employs a convergent fan of monochromatic radiation and, ideally, a 2D detector to collect the resulting fan of specularly-reflected radiation from the sample. Two variations of this idea have been previously developed. The first of these, described by \citeasnoun{naudon1}, produces a convergent beam by using a combination of a line source and a knife edge placed close to the sample surface. The presence of the knife edge permits only scattering from immediately below the knife edge to reach the detector. In effect, although the beam from the source radiates in all directions, the measurement only makes use of rays described by a fan as in Fig. \ref{recipmap}.

A second method for achieving this geometry, proposed by \citeasnoun{jmono}  and demonstrated by \citeasnoun{jmono2},  employs a Johansson monochromator (a bent monochromator in reflection geometry) to generate the convergent monochromatic radiation fan.  

Both the Naudon and Niggemeier methods have been demonstrated with lab-based sources and, in the case of the Niggemeier approach, in conjunction with a UHV chamber \cite{jmono2}. But, both methods also present challenges when considering employing them for fast and/or in-situ processes.  Clearly, requiring a knife edge in the vicinity of a sample severely limits access to that sample for other purposes, such as deposition. And, since both methods require a highly divergent monochromatic source, neither are naturally compatible with synchrotron radiation, limiting applications requiring fast time resolution.

Here, we describe a new implementation of the approach illustrated in Fig. 1 for the parallel collection of XRR data for the study of in-situ, time-resolved processes. Specifically, we utilize a polycapillary x-ray optic to create a converging fan of radiation from a monochromatic, parallel x-ray beam generated by a synchrotron source.  This allows for measurements of reflectivity curves using the framework of Naudon's method but with orders of magnitude higher flux.  This higher flux, along with a suitably fast detector, allow for a proportional decrease in collection times and improvement in time resolution.  We demonstrate continuous collection of reflectivity curves with integration times as short as 100 ms during in-situ heteroepitaxial thin film growth and resolve thickness fringes for films up to 76 nm thick at 11.4 keV.

This work was motivated, in part, by the observation of growth-induced phase transition within a complex oxide heterostructure. In particular, \citeasnoun{Ferguson} observed that growing \ce{SrTiO3} (STO) in oxygen-poor conditions on top of a stoichiometric \ce{La_{0.7}Sr_{0.3}MnO_3} (LSMO)  layer resulted in a transition of the buried LMSO layer from a Perovskite (PV) to a Brownmillerite (BM) structure. Compared to the PV structure, the BM structure is characterized by an ordered array of oxygen vacancies. Evidently, the oxygen affinity of STO is sufficient to extract oxygen from LMSO, inducing the transition. Because the transition occurs only above a certain threshold growth rate of STO, characterizing the structure and dynamics of the transition using traditional XRR was not possible. The method described here allows characterization of critical aspects of this transition, namely the volume, strain, and morphology of the BM phase during the PV-to-BM transition. The parameters for the specific implementation of qXRR described below were optimized for this particular measurement. 


The outline of the paper is as follows: after further describing the details of our approach to qXRR in \S\ref{sect:method}, we provide a theoretical framework for the design of our reflectometer and describe its theoretical resolution in \S \ref{sect:design}.  Section \S \ref{sect:exp} describes our experimental set-up and reflectivity extraction methods in detail.  Finally in \S \ref{sect:results} we show our results: characterization of the optic, a comparison of our method of qXRR with traditional Parratt XRR, and  time resolved measurements of the growth of LSMO on STO.
%
%
%

\section{Description of method} \label{sect:method}

\begin{figure}
\caption{(a) A single log-scale detector image from qXRR of a Bismuth Ruthinate thin film on Yittrium stabilized Zirocnia (111) (BRO/YSZ), showing Kiessig fringes. The reflectivity curve extracted from this image is showing in Fig. \ref{compfig}(b).  (b) and (c) show schematics of the measurement set-up.  (b) shows a top view of the beamline layout inside the hutch at the scattering plane.  (c) illustrates a side view of the entire beamline including upstream optics.  Lengths and component sizes in both views (b) and (c) are not to scale.}
\includegraphics{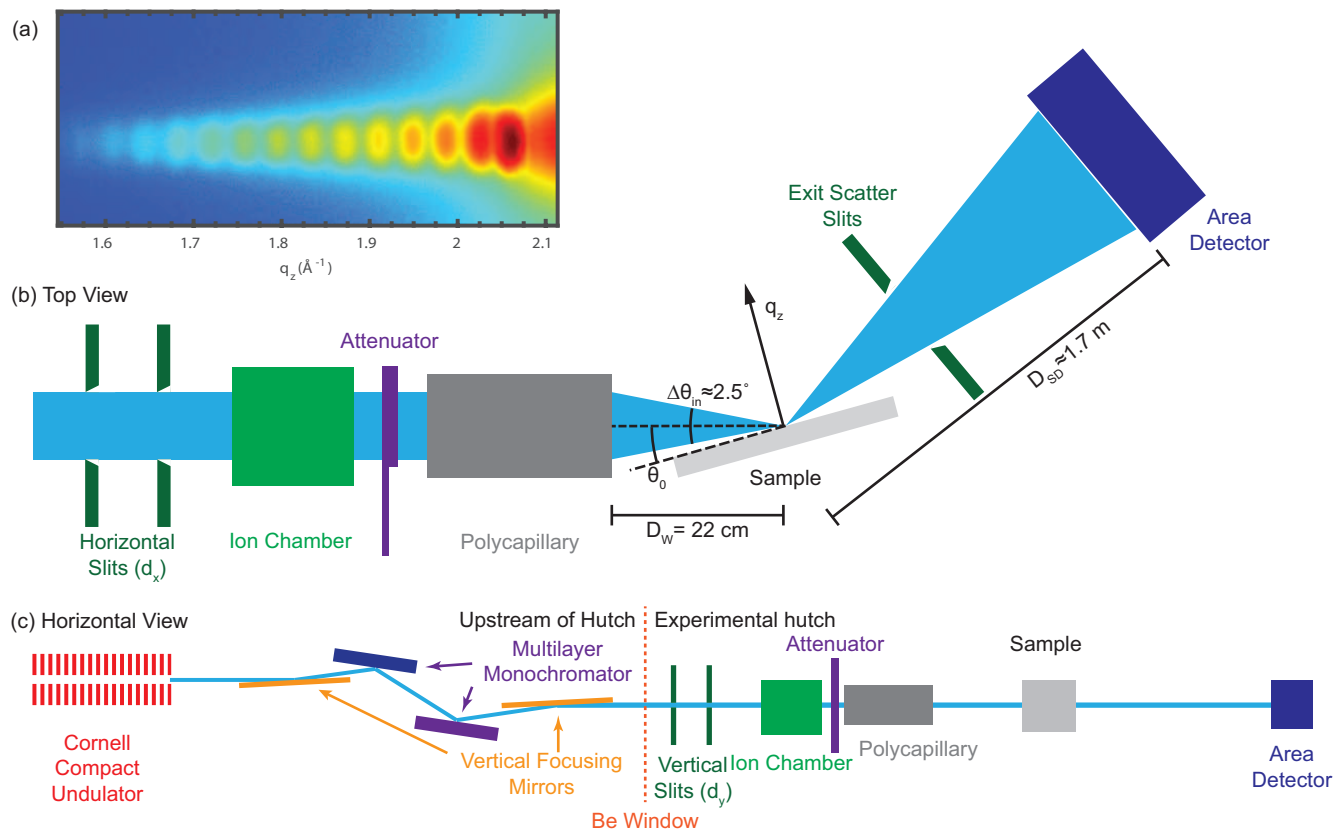}
\label{schem}
\end{figure}

The geometry of our setup is shown schematically in Fig. \ref{schem} (b) and (c).  A wide, collimated, synchrotron beam with width $d_\text{x}$ is focused on the sample surface by a polycapillary x-ray optic.  The polycapillary creates a fan of incident radiation with angular width, $\Delta\theta_\text{in}$:
\begin{equation}
\label{angrangeq}
\Delta\theta_\text{in} \approx 2\arctan\left(\frac{d_\text{x}/2}{D_\text{W}} \right)
\end{equation}
The sample is oriented such that the surface normal is in the plane of the fan.  Angle $\theta_0$, between the center of the fan and the surface, is chosen based on the portion of the reflectivity curve one wishes to collect (and such that $\theta_0>\Delta\theta_\text{in}/2$).   An area detector is placed downstream of the sample at a distance, $D_\text{W}$, determined by the desired angular resolution (see \S \ref{subsect:res}) and fan width, $\Delta\theta_\text{in}$.  
  
Fig. \ref{schem}(a) shows an image obtained by the setup described above, obtained from Bismuth Ruthinate thin film on Yittrium stabilized Zirocnia (111) (BRO/YSZ).  The \qz{}-axis is defined by the constraint that $\theta_{\text{in}}=\theta_\text{out}$. The lobes in the image are Kiessig fringes; their width in the horizontal direction in the image is determined by the thin-film thickness. The fact that the lobes are broad, rather than sharp, in the vertical direction is caused by the vertical beam-divergence introduced by the polycapillary. In the case of an ideal one-dimensional converging beam with no divergence out of plane of the fan, the vertical width of the reflectivity curve on the detector, perpendicular to the specular direction, would remain comparable to the incident beam size at the sample.  However, as expected, this width is comparable to the FWHM of the far-field of the direct beam, as discussed in \S \ref{poly_char}.


The approach described here requires relatively smooth surfaces, for which the reflected intensity is confined to a narrow region along $\hat{n}$. In this case, each pixel position along the detector parallel to the scattering plane measures intensity from a unique \qz{} value. For any real surface, there is a non-specular component of the scattered intensity that will act as a non-uniform background, as described in \S \ref{subsect:res}.
 
This approach is also compatible with a laboratory source and may provide advantages compared to prior lab-source implementations of qXRR, described in \S \ref{sect:intro}, including simplicity and cost.  Since this method uses monochromatic radiation, it can make good use of the bright characteristic line of a lab-source. 
 In this case, the optic used here would be replaced by a full-focusing optic, allowing divergent x-rays from a point source to be efficiently utilized.

\section{Theory and design} \label{sect:design}
\subsection{Polycapillary}
\label{polycap}
This method makes use of a half-focusing polycapillary x-ray optic to focus a wide, ribbon-shaped, collimated x-ray beam to a small focal spot. The details of how polycapillaries work can be found in \citeasnoun{polycap1} and elsewhere.  In short, a polycapillary optic is composed of a bundle of  small, hollow,  glass capillaries.  The bundle is heated and stretched such that, at one or both ends of the capillary bundle, the tubes are are all pointed towards a single focal point.  The bending radius of each tube is large enough such that x-rays entering a capillary within the critical angle, $\theta_\text{c}$, of its axis   
 undergo total external reflection from the glass wall and will propagate along capillary tube.   
 The critical angle of any given material is a property of the material and photon energy and defines the maximum angle for which there is total external reflection; for $\sim{}10$ keV x-rays and an air/solid interfaces $\theta_\text{c}$ is on the order of a few tenths of a degree \cite{als2011elements}.
 The efficiency of a polycapillary is the fraction of photons incident on the optic that are emitted on the downstream side.
The working distance is determined by the taper of the capillary bundle.  For an ideal capillary (one where the spot from all capillaries overlap) the beam waist, $d_\text{spot}$, at the focal point is determined by the the capillary diameter, $c$, and the amount the beam spreads out over the working distance due to the divergence, $\beta$,  from each individual capillary:
\begin{equation}
d_\text{spot}\approx\sqrt[]{c^2+(D_\text{W}\times\beta)^2}.
\label{spoteq}
\end{equation}
$\beta$ has been determined experimentally to be approximately $1.3\theta_\text{c}$ \cite{polycap1}.  Typically, the capillary diameters are negligibly small such that the spot size is approximately $D_\text{W}\times\beta$.   

\subsection{Reciprocal space resolution and its effect on background} \label{subsect:res}

\begin{figure}
\caption{Schematic diagram of qXRR, as implemented here, illustrating the regions of reciprocal space probed by three particular pixels on the detector, the diffraction geometry, and its effect on the resolution.  The probed regions for each of three pixels are shown in red, green, and blue.  Each region is bound by a set of two arcs with parallel sides corresponding to the range of incident and exit angles contributing to intensity on that pixel.  The dark features around the \qz vector represent specular features.  The wider, lighter ellipses represent less intense diffuse features in reciprocal space.}
\includegraphics{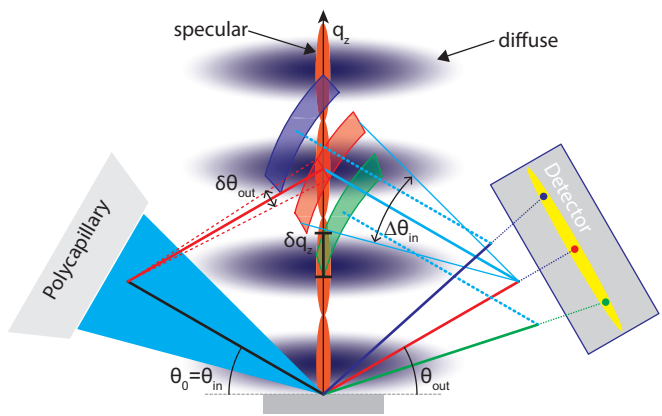}
\label{resolution}
\end{figure}

To describe the resolution of this method, we consider the region of reciprocal space over which one pixel integrates.  Fig. \ref{resolution} shows this region for each of three particular pixels.    
The long sides of each bounding box are swept out by the range of incident vectors, 
$\Delta\theta_\text{in}$ and has length $k\Delta\theta_\text{in}$  in reciprocal space.  The short sides of the area are determined by the range of exit angles that can be collected by the pixel, $\delta\theta_\text{out}$ and have length $k\delta\theta_\text{out}$.  For values consistent with our diffractometer, these arcs are very short relative to their radius and can therefore be modeled as 4 straight lines which enclose a parallelogram. 

\begin{figure}
\caption{Schematic illustration of the angular acceptance of each pixel, $\delta\theta_\text{out}$, determined as the angle between  extreme rays from the edges of the beam waist and the edges of the pixel.}
\includegraphics{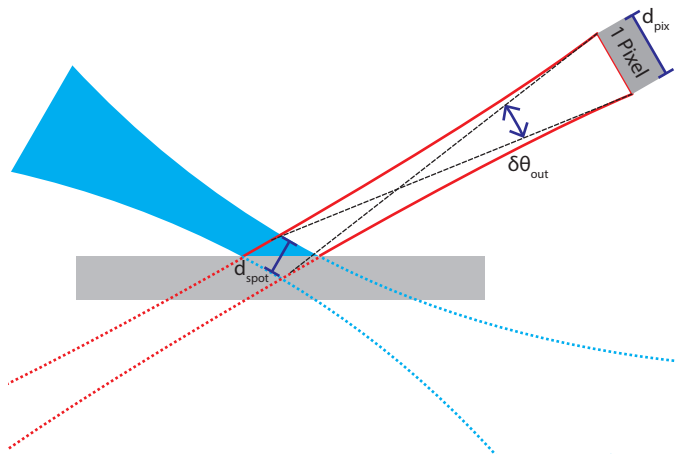}
\label{pixdiv}
\end{figure}

The angle subtended by a pixel is a function of the transverse beam size at the sample ($d_\text{spot}$), the sample to detector distance ($D_\text{SD}$), and the size of the pixel, $d_\text{pix}$, as shown schematically in Fig. \ref{pixdiv}.  Specifically:
\begin{equation}
 \delta\theta_\text{out}=2\arctan\left(\frac{d_\text{pix}+d_\text{spot}}{2D_\text{SD}}\right).
\end{equation} 

The resolution in \qz{} is given by the height of these probed regions along $\hat{n}$. Assuming a parallelogram shaped region,
\begin{equation}
\delta q_\text{z}=2k\delta\theta_\text{out}\cos(\theta_\text{out})
\label{reseq}
\end{equation}


Real samples have non-ideal features, such as correlated roughness and crystalline defects, that lead to diffuse scattering as indicated schematically in Fig. \ref{resolution}. Such scattering often extends in sheets parallel to the surface \cite{sinha,sinha1996off}, and in general can vary either in-phase or out-of-phase with the specular intensity. In Fig. \ref{resolution}, the top pixel (blue) is only seeing one Kiessig fringe along the specular.  However, that pixel is also integrating over diffuse intensity from the fringe below, which will show up as background under the reflectivity curve.  The signal to background ratio will depend on the details of the instrument and the sample: the size of $\Delta\theta_\text{in}$, the feature spacing, and the relative intensity of the diffuse and specular scattering.  As we will show in \S \ref{subsect:reflcomp}, for many samples the diffuse is weak enough to allow for the collection of usable XRR curves.  Strategies for dealing with this strong background are also discussed.

\subsection{Design considerations}

As described above, this work was motivated by an effort to monitor a Brownmillerite to Perovskite phase transition in a buried epitaxial layer of LSMO on STO.
Based on previously obtained data at 11 keV, we determined that at that energy an incident angular range of 2.1\degree{} and a resolution of 0.08\degree{} would be the minimum required to allow us to characterize the transition. 
Using equation (\ref{reseq}), we calculated that achieving this resolution required a beam waist of less than 2 mm, using a sample-to-detector distance of 1.6 m.  By using equation (\ref{angrangeq}) and (\ref{spoteq}) we determined these parameters could be met using an optic with a 22 cm working distance and an incident beam of 11 mm. This working distance was chosen to be just long enough to work with our existing vacuum system in order maximize the fan width and minimize the beam waist at the sample. With our actual design we calculate a theoretical \qz{} resolution of better than $5\times10^{-3}$ \AA$^{-1}$ for \qz$>.05$ \AA$^{-1}$.

It should be noted that the parameters described here were determined by the constraints of our experiment, especially the large working distance required by our chamber. For a given incident beam width, decreasing the working distance generally improves range of film thicknesses that can be accessed with this technique. For instance, given the same detector as used here, a setup utilizing a 5 mm-wide, 10 keV incident beam, and an optic with a 5 cm working distance would allow us to resolve Kiessig fringes from films ranging in thickness from approximately 1 nm to 80 nm, a larger range than for our actual setup.

\section{Experimental} \label{sect:exp}
\subsection{Polycapillary}

\begin{figure}

\caption{(a): an image of the direct beam from the polycapillary at the detector in log-scale. (b):  a sum of the intensity across the detector.  (c): the intensity of a vertical slice, out of the diffraction plane, at the middle of the fan.  Each pixel is 172 \um and the FWHM of the fan in the vertical direction is 5.8 mm.  The signal and background region, as described in \S \ref{subsect:reflext}, are shown.}
\includegraphics{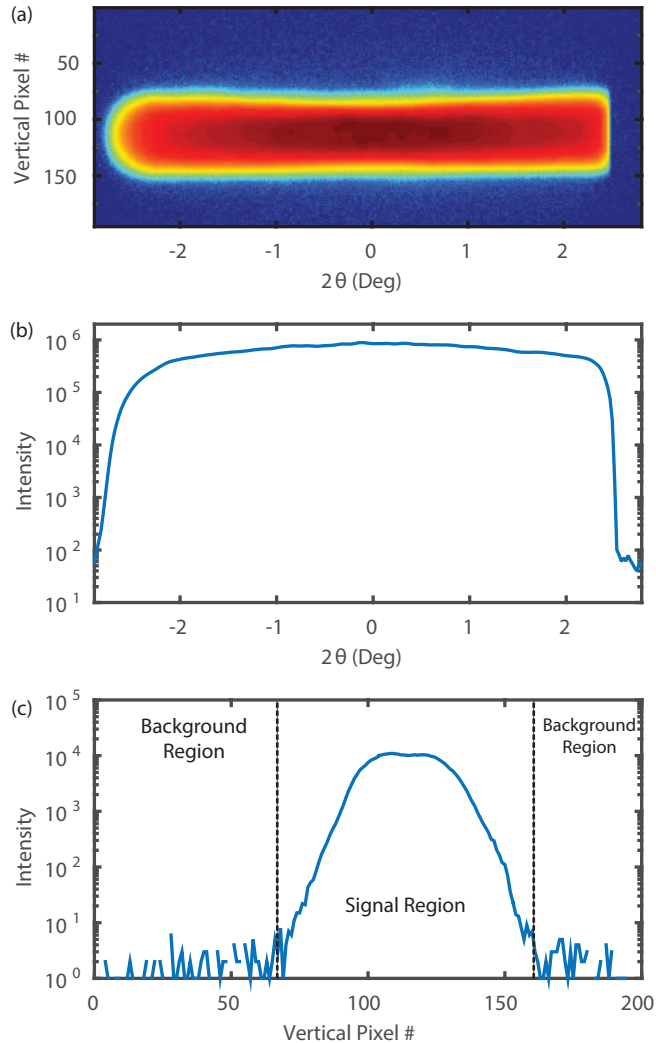}
\label{farfield}
\end{figure}
A custom half-focusing polycapillary, optimized for focusing at 11 keV, was purchased from X-ray Optical Systems, Inc (XOS, East Greenbush, NY).  The capillary has a diameter of 11 mm and working distance ($D_\text{W}$) of 22 cm from its front end.  The individual capillaries each have a 10 \textmu m diameter. The optic is encased in a vented stainless steel tube with a Be window on each end.  A removable 11$\times$1 mm slit on the upstream end indicates the most uniform portion of the optic as determined by the XOS.  Fig. \ref{farfield}(a) shows the farfield of the optic at the detector position without a sample in place. As seen in Fig. \ref{farfield}(a) and (b), the optic produces a fan of just under 2.5\degree.
\subsection{The G3 hutch at CHESS}
The measurements were performed at the G3 hutch at the Cornell High Energy Synchrotron Source (CHESS).  As shown schematically in Fig. \ref{schem}(c), the beamline is fed from a Cornell Compact Undulator type insertion device \cite{ccu}.  An internally water-cooled, Ru/\ce{B4C} double bounce multilayer monochromator (DMM) (Rigaku Innovative Technologies, Inc., Auburn Hills, MI) is used to select a beam energy of 11.4 keV with a bandwidth of $0.6\%$ $\Delta E/E$.  This energy was selected to match the fifth harmonic of the peak field of the undulator. Two mirrors, one upstream and one downstream of the DMM, provide harmonic rejection and minimize the heat bump on the multilayers.   Typically, a bendable toroidal focusing mirror is used for the downstream mirror providing a double-focused beam of $\sim1.5\times10^{14}$ photons/sec in a $0.6\times 0.6$ mm$^2$ spot. In order to produce a beam wide enough to fill the width of the capillary, in this case the toroidal mirror was replaced with a flat mirror.  Meridional focusing by bending the flat mirrors produces a  vertical spot size of  $\sim0.4$ mm at the upstream end of the capillary.
\subsection{Hutch and chamber}

At the upstream end of the G3 hutch, a set of two horizontal and vertical slits act to define the beam-size on the capillary.  These slits were set to 11 mm wide by 0.4 mm tall.  Downstream of the second slit, a \ce{N2} ion chamber serves as a proportional counter for the total flux into the polycapillary.  A variable aluminum attenuator is placed between the ion chamber and the polycapillary. The polycapillary is mounted on a stack of 4 motorized stages, two linear stages and two tilt stages,  to align the optic axis with the beam axis.  

The diffractometer consists of a motorized 3-axis table incorporating a vertical rotation axis and horizontal and vertical linear translations that are transverse to the beam.  
  A detector arm is mounted on the table which has a vertical rotation stage and a linear stage that constitute a virtual rotation about a vertical axis at the sample position.  A Pilatus 100k (Dectris LTD, Baden-Daettwil, Switzerland) detector was mounted on the arm 1.6-1.8 m from the sample.  The detector is a low-noise, photon counting detector with 195-by-487 pixels on a 172 \um pitch.  A horizontal slit was place on the detector arm, just downstream of the downstream Be window, to minimize background scatter from sources other than the sample (primarily the upstream Be window and sample holder).  A helium-filled flight-path was placed between the chamber and the detector to minimize air scatter and attenuation.  For some measurements a Si wafer was used to uniformly attenuate x-rays incident on a portion of the detector.

The table has a set of rails to allow vacuum chambers to be mounted to the table. The PLD chamber used for this experiment is described in detail by \citeasnoun{dalethesis} and has been used in a variety of in-situ experiments \cite{fleet,ferguson2009measurements,sullivan2015complex,gutierrez2015epitaxial}.
In short, the sample is mounted with its sample normal horizontal and perpendicular to the beam at $\theta_\text{0}=0$.  Two x-ray transparent Be windows allow the incident beam to reach the sample and the diffracted beam to reach the detector.
   For reflectivity scans, the table rotation stage provides the $\theta$ motion of the sample and the composite detector stage provides an angular motion that, along with the table motion, define the detector angle, $2\theta$. 

\subsection{Thin film growth}
We grew LSMO on STO by PLD.  The STO single crystal substrates were etched and annealed to produced smooth terraced surfaces with \ce{TiO2} terminations using the recipe prescribed by \citeasnoun{etch}.  Atomic Force Microscopy (AFM) was performed on each substrate after annealing to verify the formation of smooth terraces.  The films were grown from a bulk target with the desired stoichiometry.  A 248 nm KrF laser had a spot size on the target of $ \sim 2.7$ mm$^2$, and the target was ablated with a laser pulse energy of $\sim3\text{ J} /\text{cm}^2$ with a repetition rate of around 1 Hz.
The films were grown at $\sim$ 600 \degree C under $\sim10^{-3}$ torr of \ce{O2}.

\subsection{Data reduction and background subtraction}
\label{subsect:reflext}
To extract reflectivity curves from the detector images, we begin by defining a signal region and a background region, each consisting of horizontal stripes across the detector. As seen in Fig \ref{farfield}(c) the specular intensity covers approximately 90 pixels vertically on the detector; these pixels constitute the signal region.  The parts of the detector above and below the signal region are defined as the background region.  An average background for each column along the detector is determined by averaging all the pixels in that column over the background region.
This background is a combination of diffuse scattering out the the diffraction plane as well as scatter from sources other than the sample that was not blocked by the scatter slits. 
The reflectivity signal is then the sum of the background-subtracted intensity in each column over all the pixels in the signal region.  Based on the position of the detector and  $D_\text{SD}$, we determine the angular position of each column of pixels.  As the angular acceptance of each pixel is smaller than the resolution (as described in \S \ref{subsect:res}), we typically bin intensity over 4 \qz{}-points together to increase the signal to noise ratio.  If attenuation was used in front of a portion of the detector, then a correction is applied to the attenuated portion of the reflectivity curve.  It would be reasonable, particularly if trying to perform line-shape analysis, to attempt to normalize the intensity in the qXRR curves by the intensity of the far-field, shown in Fig. \ref{farfield}(b), to account for non-uniformity in the incident intensity as a function of angle. However, the presence of diffuse scattering complicates this normalization by interfering with the one-to-one correspondence between these two intensities. Therefore, we have chosen not to do this normalization. For time-resolved data we have, however, normalized the intensity of each image to the total incident flux as measured by the ion chamber upstream of the polycapillary.

\section{Results and discussion \label{sect:results}}

\subsection{Polycapillary optic characterization}
\label{poly_char}
\begin{figure}
\caption{The efficiency as a function of beam position along the upstream end of the polycapillary.  Measured by scanning a 20 \textmu m wide beam across the capillary.  The average efficiency across the middle 11 mm is shown.}
\includegraphics{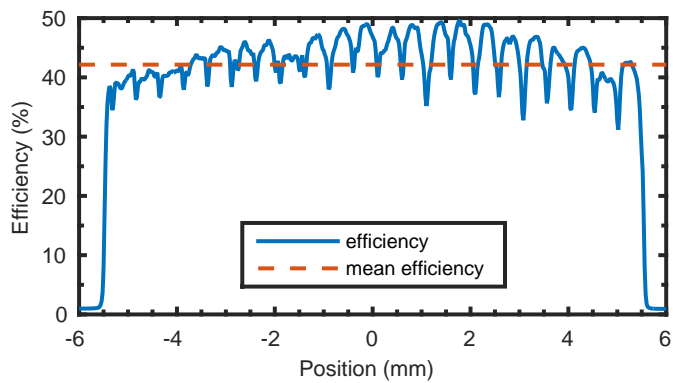}
\label{eff}
\end{figure}

The performance of the polycapillary optic used for these experiments can be characterized by three main parameters: efficiency, beam waist, and uniformity.  To measure the efficiency, we used the slits upstream of the polycapillary to define a  20 \textmu m (horizontal) $\times$ 0.4 mm (vertical) beam.  While the capillary was scanned horizontally through the beam,  \ce{N2} ion chambers upstream and downstream of the optic measured incident and transmitted intensity respectively.  Fig. \ref{eff} shows the ratio of the transmitted intensity through the capillary to the transmission with the capillary removed, as a function of position along the polycapillary.  The transmission is fairly uniform, with slight drops in transmission towards the edges due to the smaller radius of curvature of the outer capillaries.
The periodic valleys are caused by the beam hitting regularly occurring defects resulting from the modular construction of the optic \cite{gao1}.  The average  transmission efficiency across the entire width of the capillary is 42\%.

The focal spot size transverse to the beam is measured by scanning an edge through the focal spot while recording intensity downstream of the slit. 
The focal size has a FWHM of 0.6 mm.    

The uniformity of the radiation fan created by the optic is illustrated in Figs.~\ref{farfield}(a-b), showing the intensity distribution of the direct beam at the detector position with no sample in place.  The intensity as a function of angle across the detector corresponds to the intensity of each incident angle at the sample.    Interestingly, the most uniform pattern is achieved by slightly misaligning the capillary angle from that which produces the highest transmitted intensity.  There are small oscillations in intensity of around 10\% and a variation across the width of the capillary of about 25\%. 
We note that the inhomogeneity of our incident fan, while noticeable, is small compared to variations in typical XRR curves. 

As shown in Fig. \ref{farfield}(c), the vertical FWHM of the far-field is 5.8 mm tall on the detector.  As mentioned in \S \ref{sect:method} this is much larger than the height of the beam at the sample position.  This expansion of the beam is caused by the polycapillary. As described in \S \ref{polycap}, the angular divergence from a polycapillary arises from both the distribution of angles along which each individual capillary within the polycapillary bundle is directed and the beam divergence from each capillary. Since the beam is relatively narrow in the vertical direction, $d_\text{y}$=0.4mm, the latter effect dominates the divergence in the vertical direction. As described in \S\ref{polycap} this divergence, $\beta$ is of order $1.3\times\theta_\text{c} = 5.3$ mrad. Using the same approach as for calculating $d_\text{spot}$  results in an approximate size of 5.9 mm, in reasonable agreement with the results from Fig. \ref{farfield}(c).

\subsection{Reflectivity comparison}\label{subsect:reflcomp}
\begin{figure}
\caption{Three comparisons of qXRR and Parratt reflectivity: (a) and (b) low and high \qz{} region of the reflectivity curve respectively from a 15 nm \ce{Bi2Ru2O7} epitaxial pyrochlore film on a YSZ (111) substrate. (c) reflectivity from a \ce{TiO2}/\ce{TiN} multilayer on Si with 13 bilayers and a total thickness of 76 nm.  Counts are normalized to the unattenuated part of the qXRR curve.}
\includegraphics{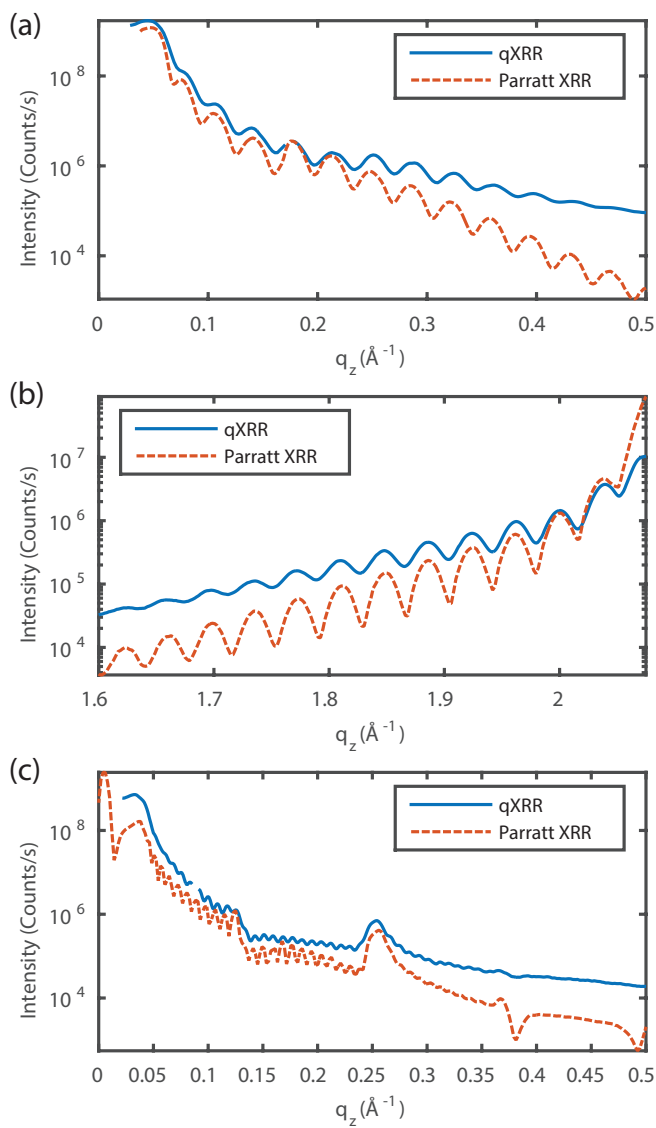}
\label{compfig}
\end{figure}

As a basic demonstration of our qXRR technique, Fig. \ref{compfig} shows three reflectivity profiles taken on static samples.  The XRR curves demonstrate both the range of angles and sample thicknesses that can be resolved.  Each pane shows the measurement taken using qXRR, with a one second exposure time, as well as a comparison scan measured by Parratt reflectivity.  The comparison scans were generated by reducing the slits to 20 $\mu m$ and performing a traditional $\theta-2\theta$ reflectivity measurement, taking several minutes to collect  (each Parratt scan has a few hundred points with 0.4 s of overhead per point, mostly associated with sample and detector motion).  Reflectivity curves were then generated by integrating the background-subtracted intensity in each detector image.   For some curves a slight shift of 2 mrad was necessary, to correct for slight mechanical errors in diffractometer alignment. 

Fig. \ref{compfig}(a) shows the reflectivity profile of a 15 nm \ce{Bi2Ru2O7} epitaxial pyrochlore film on a yittria stabilized zirconnia (YSZ) (111) substrate \cite{wakabayashi2017rotating}.  In the qXRR profile, oscillations are present through most of the range of the measurement.  At the higher \qz{} end of the curve the background becomes a large contribution to the signal and dampens the oscillations.  This background arises from diffuse scattering emanating from the sample as described in \S \ref{subsect:res} (We simulate this effect on the Parratt XRR in Fig. S2).  Because this background comes from within the scattering plane, it is not removed by our background subtraction procedure  described in \S \ref{subsect:reflext}.

Fig. \ref{compfig}(b) shows a higher angle portion of the reflectivity curve ($\theta_\text{0}=9.1\degree$) from the same sample.  This is the extracted XRR curve from the detector image shown in Fig.  \ref{schem}(a).  As above, the curves exhibit the same periodicity as compared to the Parratt reflectivity but show features that are damped  due to increased background. 

To demonstrate the resolution in \qz{} we show XRR of a thick film.  Fig. \ref{compfig}(c) shows a comparison of a \ce{TiO2}/TiN multilayer with a total thickness of 76 nm.  This figure demonstrates the range of thicknesses we can resolve.  While some of the thickness oscillations are fully or partially washed out in the qXRR, many remain visible along with the salient features of the reflectivity profile that are visible in the Parratt reflectivity.

Because the background under the qXRR curve caused by diffuse scatter varies along the specular direction and is strongest in the scattering plane, its intensity cannot be directly obtained from the image in order for it to be subtracted from the specular intensity. Rather, determining the strength of this background requires modeling it based on knowledge of the sample. This approach has been demonstrated by \citeasnoun{stoev} with data obtained by the Naudon technique. 

Although this background cannot be easily subtracted, there are strategies for minimizing its contribution.
Based on the resolution function described in \S \ref{subsect:res}, the background from off-specular scattering can be minimized by reducing the angular width of the incident beam, $\Delta\theta_\text{in}$, at the cost of the angular range of the measurement. In addition, the size of the background is significantly affected by the choice of $\theta_0$, the angle of the center of the fan. By using a $\theta_0$ such that features in the XRR curve that are associated with strong diffuse scatter are near the edge of the measured \qz{} range, the background can be significantly reduced (see Fig. S1). 
Another method, demonstrated with a tube source by \citeasnoun{Voegeli:rg5123} in a configuration similar to that developed by Niggemeier, is to put the sample normal, $\hat{n}$, 45\degree{} 
out of the plane of the incident fan.  This has the effect of separating the direction of the diffuse scatter from the specular reflection on the detector.

\subsection{Time-resolved heteroepitaxy}

\begin{figure}
\caption{Real-time reflectivity curves collected during heteroexpitaxial growth of 
\ce{La_{0.7}Sr_{0.3}MnO_3} 
on \ce{SrTiO3}.  Each trace is taken with a 0.1 s exposure time.  Only every 200th pulse is shown.  The curves are staggered by one order of magnitude for clarity.  The shoulder around the Bragg beak is attributed to background from from the diffuse scattering at the \qz{} of the Bragg peak.  Vertical tick marks are 5 orders of magnitude apart.}
\includegraphics{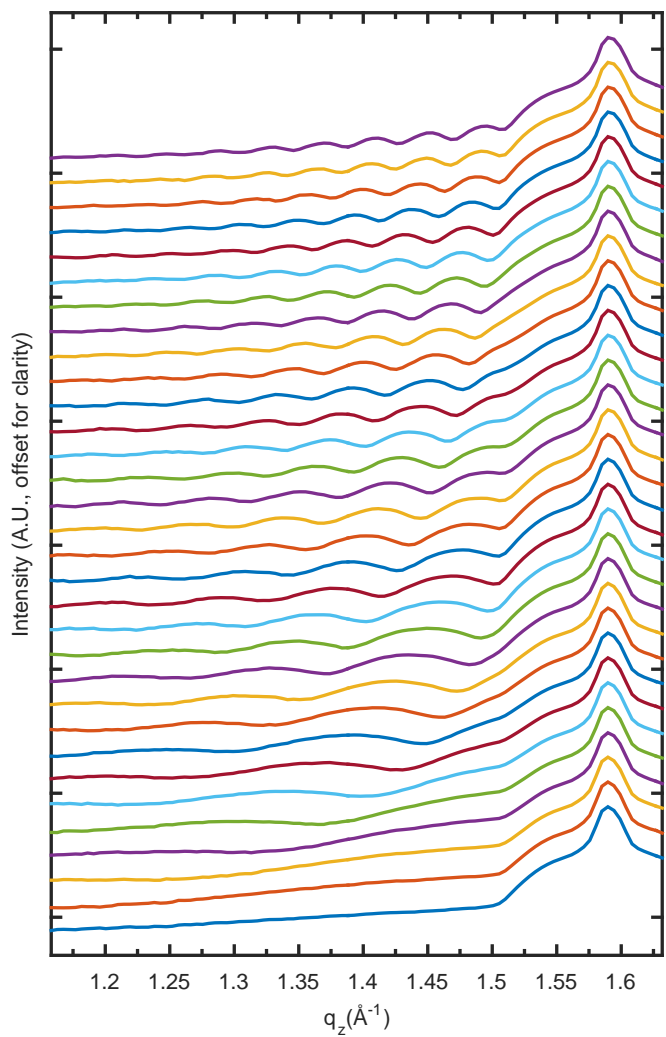}
\label{realtime}
\end{figure}

 To demonstrate the time resolved capabilities of this measurement method we collected reflectivity curves during the deposition of LSMO by pulse laser deposition. Excluding applications of qXRR, there are two ways in the literature that time-resolved XRR measurements are typically performed: either by performing Parratt reflectivity scans with low time resolution on processes that are sufficiently slow or that can be halted \cite{sinsheimer2013situ,lee_dynamic,Devloo} or by recording the intensity at one \qz{}-point as a function of time, typically near the anti-Bragg \cite{Nahm,Golks,fleet}. We show two sets of qXRR data that are analogous to each of these methods.  Analogous to the former case, in Fig. \ref{realtime} we show a portion of the reflectivity curve changing during growth. Curves from every 200th frame are shown.  At the bottom of the figure is the first curve with zero film thickness and at the top is the end of the growth with a film thickness of ~15 nm.  Curves were collected with a 0.1 s integration time (framed at ~9 Hz). 
Each curve is derived as described above.  As growth proceeds and the film thickens, the Kiessig fringes narrow and move towards higher \qz{}.  
The shoulder around the Bragg peak starting at 
$q_\text{z}=1.52 $\AA$^{-1}$ 
is not present in a pre-growth Parratt reflectivity scan and is constant in time.  We therefore ascribe it to diffuse scattering at the \qz{} of the Bragg peak, similar to that illustrated in Fig. \ref{resolution}.

\begin{figure}

\caption{Time traces for three different \qz{}-points collected during heteroexpitaxial growth of 
\ce{La_{0.7}Sr_{0.3}MnO_3} 
on \ce{SrTiO3}.  The reciprocal space location for each trace is also given in reciprocal lattice units, (00L).  Each time point is binned to 0.3 s exposure time.  The  pulse frequency of the laser was 1 Hz.}
\includegraphics{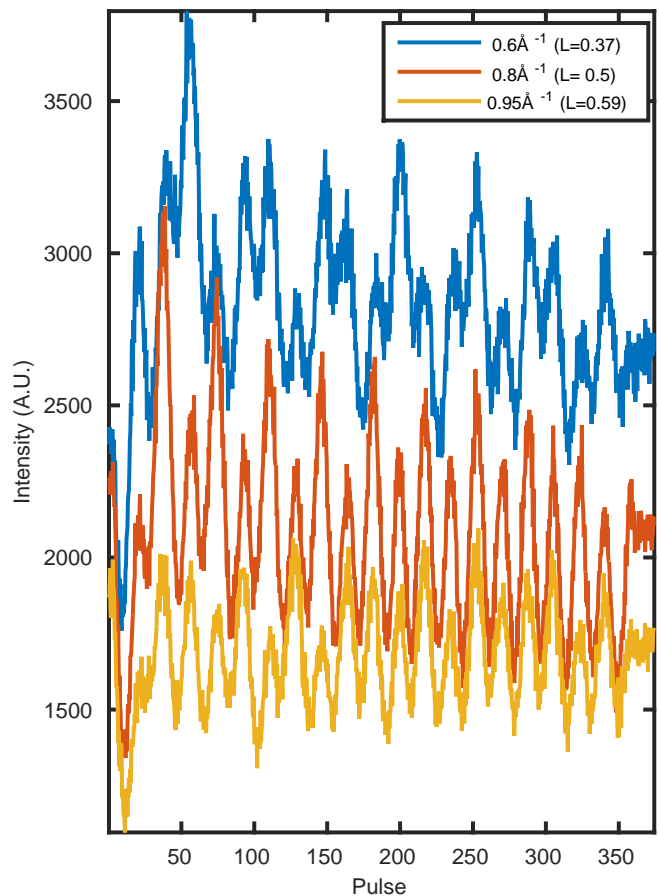}
\label{timetrace}
\end{figure}

Fig. \ref{timetrace} shows the analogous case to collecting anti-Bragg intensity as a function of time.  Using qXRR, we can measure not just the intensity at a single \qz{} -point during the deposition, but many points.
Here we show time traces for three \qz{} values, 0.6, 0.8, and 0.95 \AA$^{-1}$,
of the $\sim 100$ recorded simultaneously during a nominally identical growth to the one shown above.  These \qz{}  position in reciprocal lattice units correspond to 0.37, 0.50, and 0.95 respectively.  In agreement with diffraction models for layer by layer growth \cite{woll_modeling}, 
each trace has maxima occurring simultaneously (coinciding with layer completion) but each has different beat frequency due to their varying positions along \qz{}.  This beat frequency, relative to the layer completion rate, can be calculated as the multiplicative inverse of the distance of the measurement point from the nearest Bragg peak in reciprocal lattice units.  Using a fast Fourier transform (FFT), we determined these beat frequencies to be 0.37, 0.50 and 0.41 times the layer completion frequency for each point respectively.  Each of these beat frequencies are within 5\% of the expected value for its \qz{} position.

\section{Conclusions}
We have developed a novel technique to collect qXRR curves using high-flux monochromatic synchrotron radiation.
Compared with other approaches to qXRR, our method is simple to implement and  is compatible with both synchrotron and laboratory-based sources. 
With our particular implementation, we show that we can resolve fringes from samples as thick as 76 nm and as thin as 3 nm.
We have demonstrated the time-resolved capabilities of our measurement by collecting in-situ qXRR data during the growth of an epitaxial oxide film by pulsed laser deposition.  While we show data recorded with a 100 ms integration time, this is only limited by the signal intensity on the detector.  By measuring higher reflectivity portions of the XRR curve, reducing the \qz{}-range, and using higher flux sources we can increase the data collection rate.  In the limit of sufficient reflected intensity, such as for measurements near the critical angle, the time resolution is only limited by the frame rate of the detector.

In addition to the time-resolved capabilities of our technique, there are other applications for its use, such as scanning-probe XRR \cite{sakurai2007instrumentation}.  Because polycapillary optics can make beams with spot sizes of $\sim$10 \textmu m, this technique is a natural fit for high spatial resolution XRR measurements.   For instance, this measurement, using an optic with an appropriately small spot size, could be used for rapid measurements of variation in film thickness or density across a large area.

The implementation of qXRR described here gives rise to a non-subtractable background due to diffuse scattering. We motivated this work by describing our efforts to characterize the formation of the BM phase in LSMO.  For this experiment, measurements of the BM phase fraction, out of plane lattice parameter, and domain thickness during the transformation can be characterized by tracking the Bragg peak intensity and position and Kiessig fringe spacing respectively; these can be accurately quantified without precise fitting. Knowing how these values change over time is sufficient to answer the relevant scientific questions.
This is one example of a class of experiments for which quantitative fitting of the XRR curve is unnecessary and for which precise Parratt XRR data cannot be collected, due to the time scale on which these processes occur. For problems that do require more precise time-resolved measurements of XRR curves we have outlined several methods for reducing this background.  We are particularly interested in combining our approach with the geometry demonstrated by \citeasnoun{Voegeli:rg5123}, as described in \S\ref{subsect:reflcomp}. 









\ack{Acknowledgements}

The authors thank the Schlom research group, Cornell University, for use of their sample preparation facility and in particular H. Paik for providing the BRO film.  In addition, we thank V.D. Wheeler, Naval Research Lab, for providing the \ce{TiO2}/TiN sample.  The authors also thank Xin Huang, Cornell University, for useful discussions and code suggestions.   We thank the CHESS operations team for their support during data collection.

\ack{Funding information}

This work is based upon research conducted at and funded by the Cornell High Energy Synchrotron Source (CHESS) which is supported by the National Science Foundation and the National Institutes of Health/National Institute of General Medical Sciences under NSF award DMR-1332208.

\referencelist{}





\end{document}



\begin{figure}[h]
\centering
\includegraphics{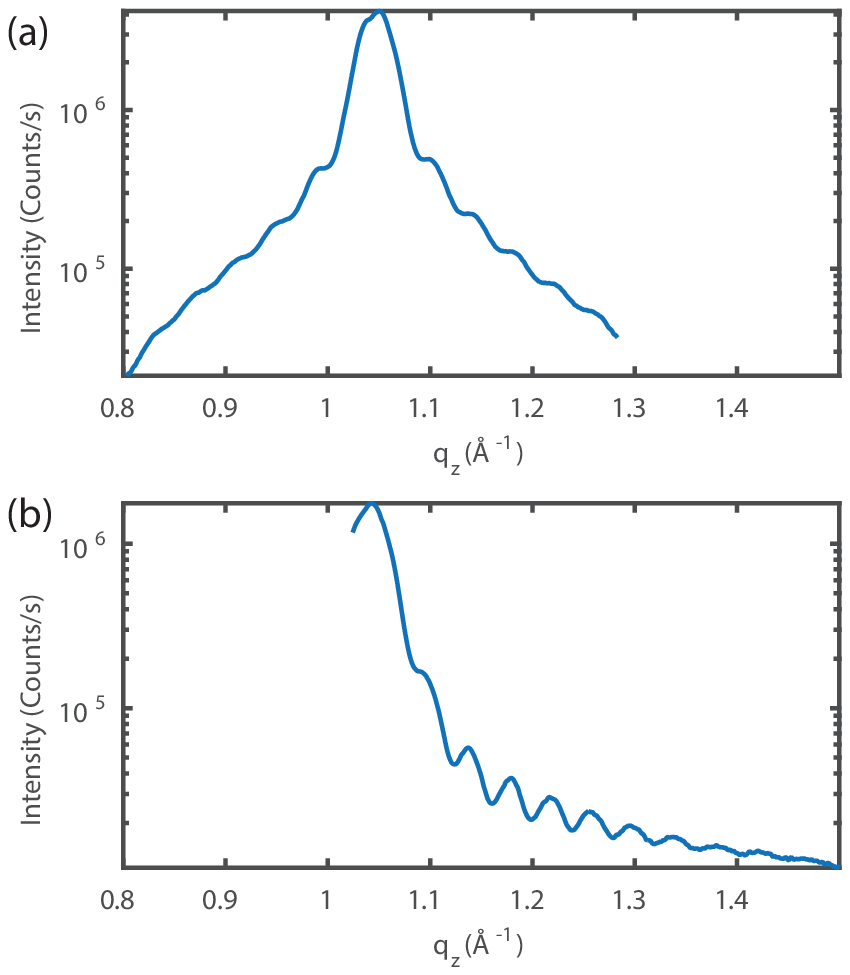}
\caption*{Fig. S1: Two different qXRR traces of BRO film on YSZ (same film as Fig. 7(a) and (b)).   
The two traces are taken with different $\theta _0$ angles.  (a) is at measured with $\theta _0=$5.18
\degree{} such that the film Bragg peak is centered in the measured range.  (b) is measured at a higher $\theta_0$ of 6.28\degree{} such that the film Bragg peak is at the edge of the measured range.  The different position of the Bragg peak along with the strong diffuse scattering extending from it parallel to the sample surface has a strong effect on the amount of background under the qXRR signal.  In (a) the diffuse at the \qz of the Bragg peak is contributing to background for the all the other \qz-values that are in the measured range and is damping out the Kiessig oscillations.  In (b), since the Bragg peak is at the edge of the measured range, the bounds of integration for most of the \qz-points in the curve do not include this strong diffuse scatter and therefore the Kiessig fringes are much stronger.
}
\end{figure}
\begin{figure}
\centering
\includegraphics{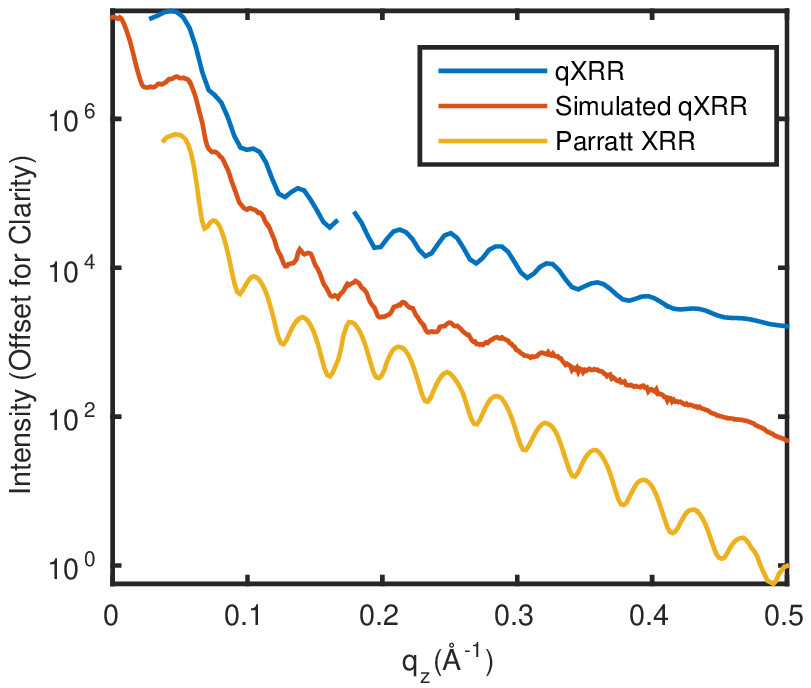}
\caption*{Fig. S2: A demonstration of the effect of diffuse scatter on the qXRR curve. We can reconstruct the qXRR data from the Parratt XRR data.  For each point in the Parratt scan we have a detector image that includes a portion of the diffuse scatter generated at each incident angle.  By offsetting and summing these images, including this diffuse background, in the correct manner, we can simulate the qXRR detector image.  Subsequently we can extract a simulated qXRR curve, shown here, that incorporates the background caused by the diffuse scatter.
}
\end{figure}